\documentstyle[twocolumn,prb,aps]{revtex}

\def\*{{\vskip3truemm}}\def\ie{{\it i.e.\ }}\def\eg{{\it e.g.\ }}
\def\V#1{{\mathbf#1}}
\def\etc{{\it etc.}}
\let\0=\noindent
\def\ciao{\end{document}}
\def\W#1{#1_{\kern-3pt\lower6.6truept\hbox to 1.1truemm
{$\widetilde{}$\hfill}}\kern2pt\,}\def\st{\scriptstyle}

\let\a=\alpha\let\b=\beta \let\g=\gamma \let\d=\delta
\let\e=\varepsilon \let\z=\zeta \let\h=\eta
\let\th=\vartheta \let\l=\lambda  \let\n=\nu
\let\x=\xi \let\p=\pi \let\r=\rho  
 \let\f=\varphi  \let\o=\omega

 \let\O=\Omega

\def\bline{\hbox to\hsize}
\newdimen\xshift \newdimen\xwidth \newdimen\yshift
\def\ins#1#2#3{\vbox to0pt{\kern-#2pt \hbox{\kern#1pt #3}\vss}\nointerlineskip}
\def\eqfig#1#2#3#4#5{\par\xwidth=#1pt \xshift=\hsize \advance\xshift
by-\xwidth \divide\xshift by 2\yshift=#2pt \divide\yshift by 2
\bline{\hglue\xshift \vbox to #2pt{\vfil#3 \includegraphics{#4.ps}
}\hfill\raise\yshift\hbox{#5}}}
\def\8{\write13}

\begin{document}

\relax
\preprint{Roma4/99/2-gg}

\title{ Quasi periodic motions from Hipparchus to Kolmogorov.}

\author{Giovanni Gallavotti}
\address{Fisica, Universit\`a di Roma ``La Sapienza''}

\date{\today}

\maketitle 
\begin{abstract} The evolution of the conception of motion as composed by
circular uniform motions is analyzed, stressing its continuity from
antiquity to our days.\cite{nota1}
\end{abstract} 
\*\*

\begin{section}{Hipparchus and Ptolemy}

Contemporary research on the problem of chaotic motions in dynamical
systems finds its roots in the Aristotelian idea, often presented as kind
of funny in high school, that motions can always be considered as
composed by circular uniform motions, \cite{[Dr53],[Ne69]}.

The reason of this conception is the perfection and simplicity of the
uniform circular motion (of which the uniform rectilinear  motion case 
must be thought as a limit case).

The idea is far more ancient than Hipparchus (from Nicea, 194-120
a.C.)  from whom, for simplicity of exposition it is convenient to
start. The first step is to understand clearly what the Greeks really
meant for motion composed by circular uniform motions.  This indeed is
by no means a vague and qualitative notion, and in Greek science it
acquired a very precise and quantitative meaning that was summarized
in all its surprising rigor and power in the {\it Almagest} of Ptolemy
($\sim$100-175 d.C.).\cite{[Ne75],[Pt84],[Ne79]}

We thus define the {\it motion composed by $n$ uniform circular motions }
with angular velocities $\o_1,\ldots,\o_n$ that is, implicitly, in use
in the {\it Almagest}, but following the terminology of contemporary
mathematics.

A motion is said {\it quasi periodic} if every coordinate of any point
of the system, observed as time $t$ varies, can be represented as:
\begin{equation}x(t)=f(\o_1t,\ldots,\o_nt)\label{(1)}\end{equation}
where $f(\f_1,\ldots,\f_n)$ is a multiperiodic function of $n$ angles,
with periods $2\p$ and $\o_1,\ldots,\o_n$ are $n$ angular velocities
that are ``rationally independent'';\cite{nota1-1} they were called
the [velocities of the] ``{\it motors}'' of the Heavens.

We must think of such function $f$ as a function of the positions
$\f_1,\ldots,\f_n$, (``{\it phases''} or ``{\it anomalies}''), of $n$
points on $n$ circles of radius $1$ and, hence, that the state of the
system is determined by the values of the $n$ angles. Therefore to say
that an observable $x$ evolves as in (\ref{(1)}) is equivalent to say
that the motion of the system simply corresponds to uniform circular
motions of the points that, varying on $n$ circles, represent the
state of the system.

We shall say, then, that the motion is {\it composed by $n$ uniform
circular motions} if it is quasi periodic in the sense of (\ref{(1)}).

In reality in Greek Astronomy it is always clear that the motion of
the solar system, conceived as a quasi periodic motion, is {\it only
one among the possible motions} of a wider family that have the form
\begin{equation}x(t)=f(\f_1+\o_1t,\ldots,\f_n+\o_nt).\label{(2)}
\end{equation}
Hence it is in a stronger sense that the motions are thought of as
composed by quasi periodic motions. Indeed all the $n$-ples
$(\f_1,\ldots,\f_n)$ of phases are considered as describing possible
states of the system. This means that one thinks that the phases
$(\f_1,\ldots,\f_n)$ provide a system of coordinates for the possible
states of the system. The observed motion is one that corresponds,
conventionally, to the initial state with phases $\f_1=\f_2=\ldots=0$;
but also the other states with arbitrary phases are possible and are
realized in correspondence of different given initial conditions and,
furthermore, {\it we can get close to them by waiting long enough}.

Summarizing: to say that the motions of a system are composed by $n$
circular uniform motions, of angular velocities $\o_1,\ldots,\o_n$ is
equivalent to say that it is possible to find a system of coordinates
that describe completely the states of the system (relevant for the
dynamical problem under study) in which the $n$ coordinates are $n$
angles and, furthermore, that in such coordinates the motion is simply
a uniform circular motion of every angle, with suitable angular
velocities $\o_1,\ldots,\o_n$.  This is indeed manifestly equivalent to
saying that an arbitrary observable of the system, evolving in the
time, admits a representation of the type (\ref{(2)}).

In Greek physics no methods were available (that we know of) for the
computation of the angle coordinates in terms of which the motion
would appear circular uniform, \ie no methods were available for the
computation of the coordinates $\f_i$ and of the functions $f$, in
terms of coordinates with direct physical meaning (\eg polar or
Cartesian coordinates of the several physical point masses of the
system). Hence Greek astronomy did consist in the hypothesis that all
the motions could have the form (\ref{(2)}) and in deriving, then, by
experimental observations the functions $f$ and the velocities $\o_i$
well suited to the description of the planets and stars motions, with
a precision that, even to our eyes (used to the screens of digital
computers), appears marvelous and almost incredible.

After Newton and the development of infinitesimal calculus it has
become natural and customary to imagine dynamical problems as
developing starting from initial conditions that can be quite
different from those of immediate interest in every particular
problem.  For example it is common to imagine solar systems in which
the radius of the orbits of Jupiter is double of what actually is or
in which the Moon is at a distance from the Earth different from the
observed one, \etc.  Situations of this kind can be included in the
Greek scheme simply by imagining that the coordinates
$\f_1,\ldots,\f_n$ are not a complete system of coordinates, and other
coordinates are needed to describe the motions of the same planets if
they are supposed to have begun their motion in situations radically
different from those which, soon or later, they would reach when
starting at the given present states (and which ``just'' correspond to
states with arbitrary values of the phase coordinates
$\f_1,\ldots,\f_n$).

To get a complete description of these ``other possible motions'' of
the system other coordinates $A_1,\ldots,A_m$ are necessary: they are,
however, {\it constant in time} on every motion and hence they only
serve to specify to which family of motions the considered one
belongs.  Obviously we shall have to think that the $\o_1,\ldots,\o_n$
themselves are functions of the $A_i$ and, in fact, it would be
convenient to take the $\o_i$ themselves as part of the coordinates
$A_i$, particularly when one can show that $m=n$ and that the $\o_i$
can be independent coordinates.

Let us imagine, therefore, that the more general motion  has the form:
\begin{equation}x(t)=f(A_1,\ldots,A_m,\o_1t+\f_1,
\ldots,\o_nt+\f_n)\label{(3)}\end{equation}
where $\o_1,\ldots,\o_n$ are functions of $A_1,\ldots,A_m$ and the
coordinates $A_1,\ldots,A_m,\f_1,\ldots,\f_n$ are a {\it complete}
system of coordinates.

In Greek astronomy there is no mention of a relation between $m$ and
$n$: probably only because no mention is made of the coordinates
$A_1,\ldots,A_m$ since the Greeks depended exclusively on actual
observations hence they could not conceive studying motions in which the
$A_1,\ldots, A_m$ (\eg the radii of the orbits of the planets, the
inclinations of the orbits, \etc ) were different from the observed
values.

In this respect it is important to remark that Newtonian mechanics
shows that it must be $m=n=3N=$ $\{$number of degrees of freedom
of the system$\}$, if $N$ is the number of bodies, even though in general
it can happen that the $\o_1,\ldots,\o_n$ cannot be taken as
coordinates in place of the $A_i$'s because they are not always
independent of each other (for instance the Newtonian theory of the
two body problem gives that the three $\o_i$ are all rational
multiples of one of them, as otherwise the motion would not be
periodic).  Nevertheless this identity between $m$ and $n$ has to be
considered one among the great successes of Newtonian mechanics.

Returning to Greek astronomy it is useful to give some example of how
one concretely proceeded to the determination of $\f_1,\ldots,\f_n$,
of $\o_1,\ldots,\o_n$  and $f$.

A good example (other than the motion of the Fixed Stars, that
is too ``trivial'', and the motion of the Sun that is, in a way, too
``simple'' to allow us to appreciate the differences between the
Ptolemaic and Copernican theories) is provided by the theory of the
Moon.

As first example I consider Hipparchus' theory of the Moon.

In general motions of heavenly bodies appear, in a first approximation,
as uniform circular motions around the center of the Earth.  Or {\it in
average} the position of the heavenly body can be deduced by imagining
it in uniform motion on a circle, ({\it ``the oblique circle that
carries the planets along''}, Dante, Par., X), with center on the Earth
and rigidly attached to the sphere (``sky'') of the Fixed Stars, that in
turn rotates uniformly around the Earth.

This average motion was called {\it deferent motion}: but the heavenly
body almost never occupies the average position, rather it is {\it
slightly} away from it sometimes overtaking it and sometimes lagging
behind.

\eqfig{200 }{120 }
{\ins{160 }{28 }{$C_0$}
\ins{72 }{110 }{$L$}
\ins{15 }{115 }{$D$}
\ins{25 }{75 }{$C$}
\ins{85 }{25 }{$\l$}
\ins{95 }{40 }{$\h$}
\ins{35 }{95 }{$\g$}
\ins{65}{5}{$O$}
}
{h1}{h1}
\*
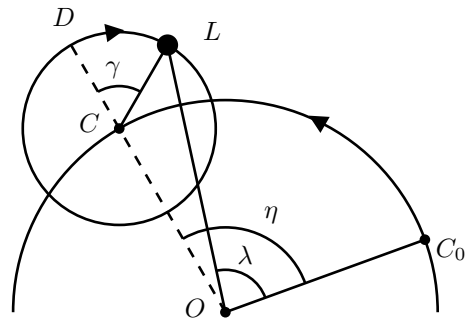
\0{\it Fig. h1}: Hipparchus' Moon theory with one deferent and one epicycle.
\*

In the case of the motion in longitude of the Moon (\ie of the
projection of the lunar motion on the plane of the ecliptic) the
simplest representation of these oscillations with respect to the
average motion is by means of two circular motions, one on a circle of
radius $CO=R$, {\it deferent}, with velocity $\o_0$ and another on a
small circle of radius $CD=r$, {\it epicycle}, with angular velocity
$\o_1$.

We reckon the angles from a conjunction between the Moon and the
average Sun, \ie when $OC_0$ projected on the ecliptic plane points at
the position of the average Sun, which is a Sun which moves exactly on
a circle with uniform motion and period one solar day (we could use
instead a conjunction between the Moon and a fixed Star, with obvious
changes).

The center of the epicycle rotates on the deferent with angular velocity
$\o_0$ and the Moon $L$ rotates on the epicycle with velocity $-\o_1$,
\cite{[Ne69]}

By using the complex numbers notation to denote a vector in the ecliptic
plane and beginning to count angles from the position of apogee on the
epicycle we see that the vector $z$ that indicates the longitudinal
position of the Moon is
\begin{equation}
\cases{\h=\o_0t\cr\g=-\o_1t\cr}\Rightarrow z=Re^{i\o_0t} +r
e^{-i(\o_1-\o_0)t}\label{(5)}\end{equation}
where $\h=\o_0t$ is the angle $C_0OC$,\ $\g=-\o_1t$ is the angle
$DCL$ and $2\p/\o_1$ is the time $T$ that elapses between two successive
returns of the Moon in apogee position on her epicycle (since
$\o_0\sim\o_1$ the new apogee will happen at a time $T$ for which there
exists a small angle $\d$ such that: $\o_0T=2\p+\d$ and
$(\o_0-\o_1)T=\d\Rightarrow$ $\o_1T-2\p=0$ $\Rightarrow T=2\p/\o_1=$
{\it anomaly month}).

In modern language we say that the Moon, having three degrees of
freedom, shall have a motion with respect to the Earth (assumed on
a circular orbit) endowed with $3$ periods: the month of anomaly (\ie
return to the apogee and ``true period of revolution'') of
approximately $27^d$, the period of rotation of the apogee of
approximately $9^y$ (in direction concording with that of the
revolution) and the period of precession (retrograde) of the node
between the lunar world and the ecliptic of approximately
$18.7^y$. This last period, obviously, does not concern the motion in
longitude which, therefore, is characterized precisely by two
fundamental periods: for instance the month of anomaly and the sideral
month (return to the same fixed star: note that the difference between
the the two angular velocities $\o_0-\o_1$ is obviously the velocity
of precession of the apogee).

Hipparchus theory of the motion in longitude of the Moon yields, as we
see, a quasi periodic motion with one deferent, one epicycle and two
frequencies (or ``motors'').  It reveals itself sufficient (if combined
with the theory of the motion in latitude, that we do not discuss here)
for the theory of the eclipses, but it provides us with ephemerides
(somewhat) incorrect when the Moon is in position of quadrature.

Ptolemy develops a more refined theory of this motion in longitude,
\cite{[Ne69],[Ne75]}.  Again assume that the angles in longitude are
reckoned from the mean Sun $S_0$ starting at a conjunction, as in the
previous theory. In a first version he imagines that the center of the
epicycle moves at the extremity of a segment of length $R-s$ that
however does not have origin on the Earth $T$ but in a point $F_1$
that moves with (angular) velocity $-\o_0$ on a small circle of radius
$s$ centered on $T$; we suppose that the center of the epicycle is
$C_1$ so that the angle between $C_1T$ and the axis of apogee (our
reference $x$--axis on a complex $z$--plane) is still $\h=\o_0t$, but
the angle $\g=-\o_1t$ that determines the position $L$ of the Moon on
the epicycle is now reckoned from $D_1$:

\eqfig{155.10000}{155}{
\ins{99.00000}{148}{$\st S_0$}
\ins{98.34000}{137}{$\st D_0$}
\ins{97.68000}{117}{$\st C_0 $}
\ins{50.82000}{100.980003}{$\st L $}
\ins{12}{93.060005}{$\st D_1 $}
\ins{118.80000}{27.720001}{$\st F_1= s e^{-i\h} $}
\ins{82}{39}{$\st F_0 $}
\ins{77}{2}{$\st T $}
\ins{86}{18.480001}{$\st \h $}
\ins{95.7}{18.48}{$\st \h $}
\ins{70}{63}{$\st R-s $}
\ins{50}{40}{$\st R'$}
\ins{105}{11.8}{$\st s $}
\ins{36.3}{88.44}{$\st \g$}
\ins{30}{74}{$\st C_1$}
}{t1}{t1}
\*
\0{\it Fig. t1}: The correction of Ptolemy to Hipparchus' theory of
the Moon.
\*

\0in formulae, if $R'=C_1T,\ \tilde R=C_1F_1=R-s$ and $\th$ is the angle 
between $D_1F_1$ and $D_0F_0$, one finds:
\begin{eqnarray}
\h&=&\o_0t,\cr
\tilde R e^{i\th}&=&R'e^{i\o_0t}-s e^{-i\o_0t}\cr
\tilde R&=&|R'e^{i\o_0t}-s e^{-i\o_0t}|,\cr
R'&\equiv&R'(\o_0t;R,s)=\cr
&=&s\cos2\o_0t+\tilde R
(1-(s/\tilde R)^2\sin^22\o_0t)^{1/2}\cr
z&=&R'(\o_0t;R,s)e^{i\o_0t} 
+r e^{-i(\o_1-\th)t}\label{(7)}
\end{eqnarray}
which reduces to Hipparchus' moon theory if $s=0$.  It also gives the
same result in conjunction and in opposition (\ie when $\h=0,\p$); it
gives a closer Moon at quadratures (\ie when $\h=\frac12\p,\frac32\p$).

This representation reveals itself sufficient for the computation of
the ephemerides also in quadrature positions, but is insufficient
(although off by little) for the computation of the ephemerides
in octagonal positions (\ie at $45^o$ from the axes).

Note that, rightly so, no new periods are introduced: the motion has
still two basic frequencies and (\ref{(7)}) only has {\it more Fourier
harmonics} with respect to (\ref{(5)}).

The theory was therefore further refined by Ptolemy himself,
\cite{[Ne69],[Ne75]}, who supposed that, in the preceding
representation, the computation of the angle $\g=-\o_1t$ on the epicycle
should be performed not by starting from the axis $C_1F_1$, as in the
previous case, but rather from the axis $F"H$ of the figure:

\eqfig{155}{185}{
\ins{99}{180}{$\st S_0$}
\ins{98.3}{169}{$\st D_0$}
\ins{97.68000}{148}{$\st C_0 $}
\ins{50.82000}{131.775604}{$\st L $}
\ins{15.84000}{125.835609}{$\st D_1 $}
\ins{118.80000}{58.515606}{$\st F_1 $}
\ins{83.82000}{69.300003}{$\st F_0 $}
\ins{92.40000}{27.495602}{$\st T $}
\ins{87.78000}{49.275604}{$\st \h $}
\ins{95.70000}{49.275604}{$\st \h $}
\ins{69.95999}{94.155609}{$\st R-s $}
\ins{104.94000}{42.675606}{$\st s $}
\ins{39.60000}{119.235611}{$\st \g$}
\ins{31.68000}{133.980011}{$\st H$}
\ins{85.80000}{11.880000}{$\st F''$}
}{t2}{t2}
\0{\it Fig. t2}: The more refined theory of Ptolemy.
\*

\0in formulae, with $R'$ as in (\ref{(7)})
\begin{equation}
z=R'e^{i\o_0t}
+r{R'e^{-i\o_0t}+se^{i\o_0t}\over|R'e^{-i\o_0t}+s
e^{i\o_0t}|}e^{-i\o_1t}\label{(9)}\end{equation}
It is clear that with corrections of this type it is possible to
obtain very general quasi periodic functions. Note that the above
theory coincides with the preceding one at conjunction, opposition and
quadratures and it is otherwise somewhat different (in particular at
the octagonal positions).

The values that Ptolemy finds for $R,r,s$, so that the theoretical
ephemerides conform with the experimental ones, are however such that
the possible variations of the Earth--Moon distance (between $R-r-s$
and $R+r+s$) are very important and incompatible with a {\it not
observed} corresponding variation of the apparent diameter of the
heavenly body. Astronomical {\it distances} (as opposed to delestial
longitudes and latitudes of planets) were not really measured in Greek
times: but we shall see that in Kepler's theory the measurability of
their value payed a major role. It is not known why the apparent
diameter of the Moon did not seem to worry Ptolemy.
\end{section}

\begin{section}{Copernicus}

Copernicus (1473-1543)\cite{[NS84]} (who was, indeed, very worried by
the latter problem) tried to find a remedy by introducing a secondary
epicycle: his model goes back to that of Hipparchus which is
``improved'' by imagining that the point of the epicycle in which
Hipparchus set the Moon was instead the center of a smaller secondary
epicycle, of radius $s$, on which, the Moon journeyed with angular
velocity $-2\o_0$

\eqfig{99.00000}{128.040009}{
\ins{21}{39.600002}{$R $}
\ins{42}{19.140001}{$\a $}
\ins{-17}{56.100002}{$C $}
\ins{-27}{90}{$\b $}
\ins{-32}{103}{$F $}
\ins{-21}{108.900002}{$\g $}
\ins{3}{106.920006}{$L$}
\ins{61}{8.580000}{$T$}
\ins{37}{117.480003}{$S_0$}
\ins{52}{87.120003}{$\cases{\a=\o_0 t&\cr
\b=-\o_1 t,\ CF=r&\cr \g=-2\o_0 t,\ FL=s&\cr}$}
}{c1}{c1}
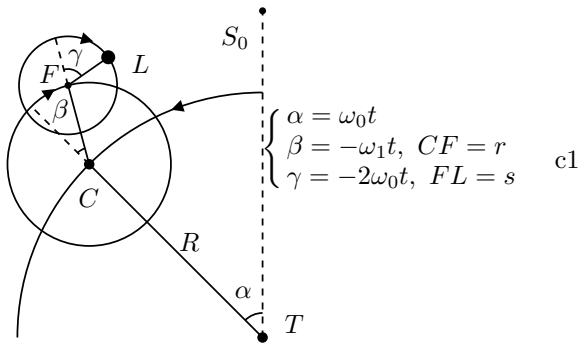
\0{\it Fig. c1}: Copernicus Moon theory with two epicycles.
\*
\0and in formulae:
\begin{equation}
r_L=Re^{i\o_0t}+e^{-(\o_1-\o_0)t}(r+s e^{2i\o_0t})\label{(11)}
\end{equation}

This gives a theory of the longitudes of the Moon essentially as
precise as that of Ptolemy. Note that, {\it again}, the same two
independent angular velocities are sufficient.

Before attempting a comparison between the method of Ptolemy and that of
Copernicus it is good to clarify the modern interpretation of the notions
of deferent and epicycle and to clarify, also, that the motions of the
Ptolemy's lunar theories are still interpretable as motions of deferents
and epicycles. Which is not completely obvious since some of the axes of
reference of Ptolemy {\it do not move of uniform circular motion}, to an
extent that by several accounts, still today, Ptolemy is ``accused'' of
having abandoned the purity of the circular uniform motions with the
utilitarian scope of obtaining agreement between the experimental data
and their theoretical representations\cite{[Sc26]}.

I just quote here Copernicus {\it Commentariolus}, few lines before the
statement of his famous second postulate setting the Earth away from
the center of the World \*

``{\it Nevertheless, what Ptolemy and several others legated to us about
such questions, although mathematically acceptable, did not seem not to
give rise to doubts and difficulties}'' ...  ``{\it So that such an
explanation did not seem sufficiently complete nor sufficiently conform
to a rational criterion}'' ... ``{\it Having realized this, I often
meditated whether, by chance, it would be possible to find a more
rational system of circles with which it would be possible to explain
every apparent diversity; circles, of course, moved on themselves with a
uniform motion''}, see\cite{[Co30]} p.108.
\*

Therefore let us check what was, in some form, probably so obvious to
Ptolemy that he did not seem to feel the necessity of justifying his
alleged {\it deviation} from the ``dogma'' of decomposability into
uniform motions. Namely we check that also the motions of the Ptolemaic
lunar theories, as actually {\it all} quasi periodic motions, can be
interpreted in terms of epicycles.

Consider for simplicity the case of quasi periodic motions with two
frequencies $\o_1,\o_2$. Then the position will be
\begin{equation}
z(t)=\sum_{\n_1,\n_2}^{-\infty,\infty}\r_{\n_1\n_2}e^{i(\o_1\n_1+\o_2\n_2)t}
\equiv\sum_j\r_je^{i\O_jt}\label{(12)}\end{equation}
by the theorem on the Fourier series, if $\n_i$ are arbitrary integers 
and if $j$ in the second sum denotes a pair $\n_1,\n_2$ and $\O_j\equiv
\o_1\n_1+\o_2\n_2$. Imagine, for simplicity, also that the enumeration 
with the label $j$ of the pairs $\n_1,\n_2$ could be made, and is made,
so that $\r_1>>\r_2\ge\r_3>\ldots$.

Then $r(t)$ can be rewritten as
\begin{eqnarray}
r(t)&=&\r_1e^{i\O_1t}\Big(1+{\r_2\over\r_1}e^{i(\O_2-\O_1)t}\cdot\bigl(\cr
&&\cdot\bigl(
1+{\r_3\over\r_2}e^{i(\O_3-\O_2)t}(1+\ldots)\bigr)\Big)\label{(13)}
\end{eqnarray}
which, neglecting $\r_2,\r_3\ldots$, is the uniform circular motion on
the deferent of radius $|\r_1|$ with velocity angular $\O_1$; neglecting
only $\r_3,\r_4,\ldots$ it is a motion with a deferent of radius $|\r_1|$
rotating at velocity $\O_1$ on which rests an epicycle of radius $|\r_2|$
on which the planet rotates at velocity $\O_2-\O_1$; neglecting only
$\r_j, \,j\ge4$ one obtains a motion with one deferent and two epicycles,
as that used by Copernicus in the above lunar model.

If $|\r_1|$ is not much larger than the other radii (and precisely if
$|\r_1|$ not is larger than the sum of the other $|\r_j|$, a situation
that is not met in the ancient astronomy), what said remains true
except that the notion of deferent is no longer meaningful. Or, in
other words, the distinction between {\it main} circular motion and
epicycles is no longer so clear from a physical and geometrical
viewpoint. The epicycle with radius larger than the sum of the radii
of the other epicycles, if existent, essentially determines the
average motion and is given the privileged name of ``deferent''. In
the other cases, although the average motion still makes sense,
\cite{[St69]}, it is no longer associated with a particular epicycle,
but all of them concur to define it, for an example see [AA68], p.138.

We see, therefore, the {\it complete equivalence} between the
representation of the quasi periodic motions by means of a Fourier
transform and that in terms of epicycles, \cite{[Sc26]}.

Greek astronomy, thus, consisted in the search of the Fourier
coefficients of the quasi periodic motions of the heavenly bodies
representing them geometrically by means of uniform motions.

But Ptolemy's method is in a certain sense not systematic (see,
however, below): the intricate interplay of rotating sticks that
explains, or better parameterizes, the motion of the Moon is very
clever and precise but it seems quite clearly not apt for obvious
extensions to the cases of other planets and heavenly bodies.

Copernicus' idea, {\it instead}, of introducing epicycles of epicycles,
as many as needed to an accurate representation of the motion, is
systematic and, as seen above, coincides with the computation of the
Fourier transform of the motion coordinates with coefficients ordered by
decreasing absolute value. Copernicus' work (with the only exception,
and such only in a rather restricted sense that it is not possible to
discuss here, of some details of the motion of Mercury) is strictly
coherent with this principle. set in his early project quoted above.

This is perhaps\cite{nota3a} the great innovation of Copernicus and
not, certainly, the one he is always credited for, \ie having referred
the motions to the (average) Sun rather than to the Earth: that is a
trivial change of coordinates, known as possible and already studied
in antiquity, \cite{[He81],[He91]}, by Aristarchus (of Samos, 310-235
a.C.), Ptolemy \etc, but set aside by Ptolemy for obvious reasons of
convenience, because in the end it is from Earth that we observe the
heavens (so that still today many ephemerides are referred to the
Earth and not to an improbable observer on the Sun), and also because
he seemed to lack an understanding of the principle of inertia (as we
would say in modern language). See the {\it Almagest}, p.45 where
allegedly Ptolemy says: ``{\it ...although there is perhaps nothing in
the celestial phenomena which would count against that hypothesis
{\rm[that the Sun is the center of the World]}... one can see that
such a notion is quite ridiculous}.\cite{nota3}

Ptolemy, with clever and audacious geometric constructions does not
compute coefficient after coefficient the first few terms of a Fourier
transform of the motion. He {\it sees} directly series which contain
infinitely many Fourier coefficients (see $R'(\o_0t)$ in (\ref{(7)})
where this happens because of the square root), \ie infinitely many
epicycles, {\it most of which are obviously very
small and hence irrelevant}. 

We can therefore obtain the same results with several arrangements of
sticks, provided that the motion that results has Fourier
coefficients, I mean those which are not negligible, equal or close to
those of the motion that one wants to represent: it is this absence of
uniqueness that makes the method Ptolemaic appear not systematic.

It has, however, the advantage that, if applied by an astronomer like
Ptolemy, it requires apparently, at equal approximation, less
elementary uniform circular motions, a fact that was erroneously
interpreted as meaning {\it less epicycles} (which, on the contrary,
are very often, in the Ptolemaic constructions, infinitely many as we
see in the case of (\ref{(7)}),(\ref{(9)})) than usually necessary
with the methods of Copernicus: a fact that was and still is
considered a grave defect of the Copernican theory compared to the
Ptolemaic. Ptolemy identifies $43$ fundamental uniform circular
motions (that combine to give rise to quasi periodic functions endowed
with {\it infinitely many harmonics} formed with the $43$ fundamental
frequencies) to explain the whole system of the World: Copernicus
hopes initially (in the {\it Commentariolus}) to be able to explain
everything with $34$ harmonics, only to find out in the {\it De
Revolutionibus} that he is forced to introduce several more. See
Neugebauer in \cite{[Ne75]}, vol. 2, p.925- 926.\cite{nota2}

One should not, however, miss stressing also that Copernicus
heliocentric assumption made possible a simple and unambiguous
computation of the planetary distances.\cite{nota3a} If the Sun is
assumed as the center, and the orbits are supposed circular (to make
this remark simplest) then the radii of the epicycles of the external
planets (for instance) are automatically fixed to be all equal to the
distance Earth--Sun. Then, knowing the periods of revolution and
observing one opposition (to the Sun) of a planet and one position off
conjunction at a later time, one easily deduces the distance of the
planet to the Earth and to the Sun, in units of the Earth--Sun
distance. In a geocentric system the radii of the epicycles are simply
related to their deferents sizes and the latter are {\it a priori}
unrelated to the Sun--Earth distance: for this reason in ancient
astronomy the size of the planetary distances was a big open
problem. One can ``save the phenomena'' by {\it arbitrarily scaling}
deferent and epicycles radii {\it independently for each planet!}  The
possibility of reliably measuring the distances, applied by Copernicus
and then by Tycho and Kepler, was essential to Kepler who could thus
see that the saving of the phenomena in longitudinal observation was
not the same as saving them in the radial observations, a more
difficult but very illuminating task, see\cite{nota-4}.
\end{section}

\begin{section}{Kepler}

Today we would say that Ptolemy's theory was {\it nonperturbative}
because it immediately represented the motions as quasi periodic
functions (with infinitely many Fourier coefficients).  Copernicus'
is, instead, {\it pertubative} and it systematically generates
representations of the motions by means of developments with a finite
number of harmonics constructed by adding new {\it pure} harmonics,
one after the other, with the purpose of improving the agreement with
experience. The larger number of harmonics in Copernicus is simply
explained because, from his point of view, harmonics multiple of
others count as different epicycles, while in Ptolemy the geometric
constructions associated with an epicycle sometimes introduce also
harmonics that are multiples, or combinations with integer
coefficients, of others already existent and produce an ``apparent''
saving of epicycles.

But the systematic nature of the Copernican method permitted to his
successors to organize the large amount of new data of the astronomers
of the Renaissance and of the Reform time. Eventually it allowed
Kepler (1571-1630) to recognize that what was being painfully
constructed, coefficient after coefficient, was in the simplest cases
just the Fourier series of a motion that developed on an ellipse with
the Sun, or the Earth in the case of the Moon, in the focus and with
constant area velocity.

For reasons that escape me the History of Science usually credits Kepler
to have made possible the {\it rejection} of the scheme of
representation in terms of deferents and epicycles, in favor of motions
on ellipses.

But it is instead clear that the Keplerian motions are still
interpretable in terms of epicycles whose amplitudes and positions are
computed with the Copernican or Ptolemaic methods (that he regarded as
equivalent in a sense that reminds us of the modern theories of
``equivalent ensembles'' in statistical mechanics, see\cite{[Ke09]}
Ch. 1-4) or, equivalently, via the modern Fourier transform.  Nor it
should appear as making a difference that the epicycles are, strictly
speaking, infinitely many, (even though all except a small number have
amplitudes, \ie radii, which are completely negligible): already the
Ptolemaic motions, with the their audacious constructions based on
rotating sticks did require, to be representable by epicycles (\ie by
Fourier series) infinitely many coefficients (or {\it harmonics}), see
(\ref{(7)}),(\ref{(9)}) above, in which the r.h.s. manifestly have
infinitely many nonvanishing harmonics.

Only Copernican astronomy was built to have a finite number of
epicycles: but their number had to be ever increasing with the
increase of the precision of the approximations. Ptolemy {\it seemed
looking} and Kepler certainly {\it was looking} for exact theories,
Copernicus appears to our eyes doomed to look for better and better
approximations.

In reality also the critique of lack of a systematic method in
Ptolemy, the starting point of the Copernican theory, should be
reconsidered and subject to scrutiny: indeed we do not know the
theoretical foundations on which Ptolemy based the {\it Almagest} nor
through which deductions he arrived at the idea of the {\it equant}
and to other marvelous devices. One can even dare the hypothesis that
the {\it Almagest} was just a volume of commented tables based on
principles so well known to not even deserve being mentioned. It is
difficult to imagine that Ptolemy had proceeded in an absolutely
empirical manner in the invention of {\it anomalous} objects like
``equant points'' and strange epicycles (like those he uses in the
theory of the Moon) and he did not feel that he was departing from the
main stream based on the axiom that all motions were decomposable into
uniform circular motions: it is attractive, instead, to think that he
did not feel, by any means, to have violated the law of the
composition of motions by circular uniform ones.

One should note that if a scientist of the stature of Copernicus in a
1000 years from now, after mankind recovered from some great disaster,
found a copy the American Astronomical Almanac\cite{[AA89]} (possibly
translated from translations into some new languages) he would be
astonished by the amount of details, and by the data correctness,
described there and he would be left wandering how all that had been
compiled: because it is {\it very difficult, if not impossible} to
derive even the Kepler's laws, directly from it (not to mention the
present knowledge on the three body problem). And he would say
``surely there must be a simpler way to represent the motions of the
planets, stars and galaxies'', and the whole process might start anew,
only to end his life (as Copernicus in ``{\it De
revolutionibus}'')\cite{[Co30]} with new tables that coincided with an
appropriately updated version of the ones he found in his youth.  The
American Astronomical Almanac can be perhaps better compared to
Ptolemy's {\it Planetary Hypotheses} if the latter is really due to
him, as universally acccepted, while the Almagest is an earlier but
more detailed version of it\cite{nota3}.

After the discovery of the Kepler laws the theory of gravitation of
Newton (1642-1727) was soon reached, \cite{[Ne62]}.  Contrary to what
at times is said, far from marking the end of the grandiose Greek
conception of motion as composed by circular uniform motions,
Newtonian mechanics has been, instead, its most brilliant
confirmation.

For example, if $\th$ denotes the angle between the major semiaxis and
the actual position on the orbit (``{\it true anomaly}''), $\ell$
denotes the {\it average anomaly}, $a$ is the major semiaxis of the
ellipse and $e$ is its {\it eccentricity}, the Keplerian motion of
the Mars around the Sun is described by the equations:
\begin{eqnarray}
z&=&p e^{i\th}(1-e\cos\th)^{-1},\ 
p=a(1-e^2)\cr
\th&=&\ell-2e\sin \ell+\frac54 e^2\sin2\ell+O(e^3),\ 
\ell=\o t\label{(14)}\end{eqnarray}
hence
\begin{eqnarray}
z&=&p(1-e^2)^{1/2} e^{i\th}(1+2\sum_{n=1}^\infty\h(e)^n\cos
n\th)\nonumber\\
\h(e)&\equiv&(1-(1-e^2)^{1/2})e^{-1}=\frac12e+O(e^3)\label{(15)}
\end{eqnarray}
and to first order in $e$:
\begin{eqnarray}
z&=& a e^{i\o t}(1-2e\sin\o t)(1+2e\cos t)+O(e^2)=\nonumber\\
&=&a e^{i\o t}(1+e(1+i)e^{i\o t}+e(1-i)e^{-i\o t}+
O(e^2))\label{(16)}
\end{eqnarray}
which can be described to lowest order in $e$, as composed by a
deferent and two epicycles. Two more would be necessary to obtain an
error of $O(e^3)$.

In this respect it is interesting to observe how one can arrive to an
ellipse with focus on the Sun, by considering epicyclical
motions. Indeed the simplest epicyclical motion is perhaps that in
which one considers infinitely many pairs of epicycles, run with
respective angular velocity $\pm n\o$, with $n=1,2,\ldots$, and with
radii decreasing in geometric progression, \ie:
\begin{equation}
z(\th)=p' e^{i \th}\sum_{n=0}^\infty\h^{\prime n}\cos n\th\label{(17)}
\end{equation}
for some $p',\h'$, that leads to the ellipse in the first of the
(\ref{(15)}).

What is {\it less natural} in the Kepler laws, is that the time law
which gives the motion on the ellipse, instead of $\th\to\o t$, is
rather $\th\to\o t-2e\sin\o t+\ldots$. Such motion is however an old
``Ptolemaic knowledge'' being, at least at lowest order in $e$, a
uniform {\it angular} motion around a point $S_{equant}$ of abscissa
$2e a$ from the point $S$ with respect to which the anomaly $\th$ is
evaluated and of abscissa $e a$ with respect to the center $C$ of the
circle on which the (manifestly nonuniform) motion takes place 

\vskip3pt

\eqfig{152}{100}{
\ins{117}{90}{$\st M $}
\ins{16}{3}{$\st S $}
\ins{49}{3}{$\st C $}
\ins{36}{23}{$\st \th $}
\ins{97}{23}{$\st \ell $}
\ins{78}{3}{$\st S_{equant} $}
\ins{60}{19}{$\st e a$}
\ins{0}{61}{$\st CM=a,\ SC= e a$}
}
{e}{e}
\* 
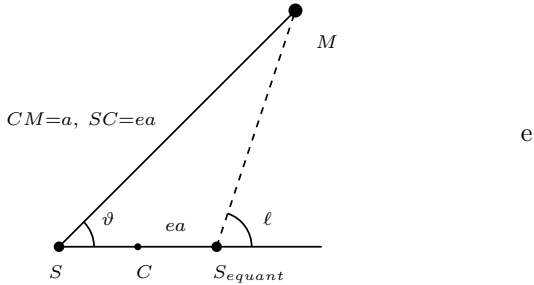
\0{\it Fig. e}: The equant construction of Ptolemy adapted to a
heliocentric theory of Mars; $S$ is the Sun, $M$ is Mars, $C$ the
center of the orbit and the equant point is $S_{equant}$.
\*

\0this means that the angle $\ell$ in the drawing rotates uniformly
and
\begin{equation}
\th=\ell-2e\sin \ell+ e^2 \sin 2\ell+O(e^3)\label{(19)}\end{equation}
Truncating the series in (\ref{(17)}) and (\ref{(19)}) to first order
in the eccentricity we obtain (\ref{(16)}) and hence a description in
terms of one deferent, two epicycles and an equant: it is a
description quite accurate of the motion of Mars with respect to the
Fixed Stars Sky and it is the theory that one finds in the {\it
Almagest}, after converting it to the inertial frame of reference
fixed with the Sun.  

The motion of the Earth around the Sun (or viceversa if one prefers)
is similar except that the center of the deferent circle is directly
the equant point, see\cite{[Ne69]} p.192, see also\cite{[Ke09]}
Ch.2-4: this is usually quoted by saying the ``for the Earth Ptolemy
(Copernicus and Tycho) did not {\it bisect} the eccentricity'',
meaning that the center and the equant were identical and both $2\,e\,a$
away from the Sun: from\cite{nota-4} we deduce that this did not
matter for the Earth which has a much smaller eccentricity (than
Mars). Before discovering the ellipse Kepler had to redress this
``anomaly'' and he indeed bisected also the Earth eccentricity,
see\cite{nota-4}, making the Copernican Earth lose one more
distinguishing feature with respect to the other planets.\cite{[nota-Co]}

The above, however, {\it is not} the path followed by Kepler,
see\cite{nota-4} where the latter is discussed in some detail.

Thus bringing the development in $e$ to first order one reaches a
level of approximation quite satisfactory for the observations to which
Kepler had access, not only for the Sun but also for the more anomalous
planets like Mercury, Moon and Mars: to second order however the equant
becomes insufficient and Kepler realized that the ellipse had to be
described at constant area velocity with respect to the focus.

We can say that {\it the experimental data agree within a third order
error in the eccentricity with the hypothesis of an elliptical motion
and with a time law based on the area law: this, within a second order
error in the eccentricity, coincides with the Ptolemaic law of the
equant}.
\end{section}

\begin{section}{Modern times}

To realize better the originality of the Newtonian theory we must
observe that in the approximations in which Kepler worked it was
evident that the laws of Kepler were not absolutely valid: the
precession of the lunar node, of the lunar perigee and of the Earth
itself did require, to be explained, new epicycles: in a certain sense
the Keplerian ellipses became ``deferent'' motions that, if run with
the law of the areas, did permit us to avoid the use of equants and of
other Ptolemaic ``tricks''. The theory of Newtonian gravitation
follows after the abstraction made by Newton according to which the
laws of Kepler, manifestly in contrast with certain elementary
astronomical observations unless combined with suitable constructions
of epicycles as Kepler himself realized and applied to the theory of
the Moon,\cite{[St94]} were {\it rigorously exact} in the situation in
which we {\it could neglect the perturbations due to the other
planets}, \ie if we consider the ``two body problem'' originally
reinterpreting the Keplerian conception that the motion of a planet
was due mostly to a force due to the Sun and partly to a force due to
itself.

The theory of gravitation not only predicts that the motions of the
heavenly bodies are quasi periodic, apparently even in the
approximation in which one does not neglect the reciprocal
interactions between the planets, but it {\it gives us the algorithms}
for computing the functions $f(\f_1,\ldots,\f_n)$.

The {\it summa} of Laplace (1749-1827) on the {\it M\'ecanique
cel\`este} of 1799, \cite{[La66]}, makes us see how the description of
the solar system motions, also taking account of the interactions
between the planets, could be made in terms quasi periodic functions.
The Newtonian mechanics allows us to compute approximately the $3N$
coordinates $\V A=(A_1,\ldots,A_{3N})$ and the $3N$ angles
$\f_1,\ldots,\f_{3N}$ and the $3N$ angular velocities $\o_1(\V
A),\ldots,\o_{3N}(\V A)$ in terms of which the motion simply consists
of $3N$ uniform rotations of the $3N$ angles while the $\V A$ remain
constant.

Laplace makes us see that there is an algorithm that allows us to
compute the $A_i,\o_i,\f_i$ by successive approximations in a series
of powers in several parameters (ratios of masses of heavenly bodies,
eccentricities, ratios of the planets radii to their orbits radii \etc),
that will be denoted here with the only symbol $\e$, for simplicity.

After Laplace approximately 80 years elapse during which the technique
and the algorithms for the construction of the {\it heavenly} series
are developed and refined leading to the construction of the formal
structure of analytic mechanics. And Poincar\'e hws able to see
clearly the new phenomenon that marks the first true and definitive
blow to the Greek conception of motion: with a simple proof,
celebrated but somehow little known, he showed that the algorithms
that had obtained so many successes in the astronomy of the 1800's
were in general {\it nonconvergent} algorithms, \cite{[Po87]}.

Few did realize the depth and the revolutionary character of
Poincar\'e's discovery: among them Fermi that tried of deduce from the
method of Poincar\'e the proof of the ergodicity of the motions of
Hamiltonian dynamical systems that were not too special.  The proof of
Fermi, very instructive and witty although strictly speaking not
conclusive from a physical viewpoint, remained one of the few attempts
made in the first sixty years of the 1900's by theoretical physicists,
to understand the importance of of Poincar\'e's theorem.

Fermi himself, at the end of his history, came back on the subject to
reach conclusions very different from the ones of his youth (with the
celebrated numerical experiment of Fermi-Pasta-Ulam).\cite{[FPU55]}

And the Greek conception of motion finds one of its last (quite
improbable, {\it a priori}) ``advocates'' in Landau that, still in the
1950's, proposes it as a base of his theory of turbulence in
fluids.\cite{[LL71]} His conception that has been criticized by Ruelle
and Takens (and apparently by others)\cite{[RT71a],[RT71b]} on the
base of the ideas that, at the root, went back to Poincar\'e.

The alternative proposed by them began the modern research on the
theory of the development of the turbulence and the renewed attempts
at the theory of developed turbulence.

The attitude was quite different among the mathematicians who, with
Birkhoff, Hopf, Siegel in particular, started from Poincar\'e to
begin the construction of the corpus that is today called the {\it
theory of chaos}.

But only around the middle of the 1950's it has been possible to solve
the paradox consisting in the dichotomy generated by Poincar\'e:
\*

{(1) } on the one hand the successes of classical astronomy based on
Newtonian mechanics and the perturbation theory of Laplace, Lagrange,
\etc\ seemed to confirm the validity of the quasi periodic conception
of motions (recall for instance Laplace's theory of the World, or 
Gauss' ``rediscovery'' of Ceres,\cite{[Ga71]} and the discovery of
Neptune, \cite{[Gr79]}).

{(2) } on the other hand the theorem of Poincar\'e excluded the
convergence of the series used in (1)).
\*

The fundamental new contribution came from
Kol\-mo\-go\-rov,\cite{[Ko54],[Ga84]}: he stressed the existence of
two ways of performing perturbation theory, \cite{[Ga71]}.  In the
first way, the classical one, one fixes the initial data and lets them
evolve with the equations of motion.  Such equations, in all
applications, depend by several small parameters (ratios of masses,
\etc) denoted above generically by $\e$.  And for $\e=0$ the equations
can be solved exactly and explicitly, because they reduce to a
Newtonian problem of two bodies or, in not {\it heavenly} problems, to
other {\it integrable} systems.  One then tries to show that the
perturbed motion, with $\e\ne0$, is still quasi periodic, simply by
trying to compute the periodic functions $f$ that should represent the
motion with the given initial data (and the corresponding phases
$\f_i$, angular velocities $\o_i$, and the constants of motion $A_i$)
by means of power series in $\e$.  Such series, however, do not
converge or sometimes even contain divergent terms, deprived of
meaning, see\cite{[Ga84]} Sec. 5.10.

A second approach consists in fixing, instead of given initial data
(note that it is in any case illusory to imagine knowing them
exactly), the angular velocities (or {\it frequencies})
$\o_1,\ldots,\o_n$ of the quasi periodic motions that one wants to
find.  Then it is often possible to construct by means of power series
in $\e$ the functions $f$ and the variables $\V A,\V \f$, in terms of
which one can represent quasi periodic motions, with the prefixed
frequencies.

In other words, and making an example, we ask the possible question:
given the system Sun, Earth, Jupiter and imagining for simplicity the
Sun fixed and Jupiter on a Keplerian orbit around it, is it or not
possible that {\it in eternity} (or also only up to just a few
billion years) the Earth evolves with a period of rotation around to
Sun of about $1$ year, of revolution around its axis of about $1$ day,
of precession around the heavenly poles of about $25.500$ years,
\etc?

One shall remark that this second type of question is much more similar to the
ones that the Greek astronomers asked themselves when trying to deduce
from the periods of the several motions that animated a heavenly body the
equations of the corresponding quasi periodic motion.

The answer of Kolmogorov is that if $\o_1,\ldots,\o_n$ are the $n$
angular velocities of the motion of which we investigate the existence
it will happen that for the {\it most part} of the choices of the $\o_i$
there actually exists a quasi periodic motion with such frequencies and
its equations can be constructed by means of a power series in $\e$,
{\it convergent} for $\e$ small.\cite{[Ko54]}

The set of the initial data that generate quasi periodic motions has a
complement of measure that tends to zero as $\e\to0$, in every bounded
part of the phase space contained in a region in which the
unperturbed motions are already quasi periodic.

One cannot say, therefore, whether a preassigned initial datum actually
undergoes a quasi periodic motion, but one can say that near it there
are initial data that generate quasi periodic motions. And the closer
the smaller $\e$ is.

By the theorem of continuity of solutions of equations of motion
with respect to variations of initial data it follows that every
motion can be simulated, for a long time as $\e\to0$, by a quasi
periodic motion.

But obviously there remains the problem:
\*

{(1) } are there, really, initial data which follow motions that, in
the long run, reveal themselves to be  not quasi periodic?

{(2) } if yes, is it possible that in the long run the motion of a system 
differs substantially from that of the (abundant) quasi periodic motions
that develop starting with initial data near it?
\*

The answer to these questions is affirmative: in many systems motions
that are not quasi periodic do exist and become easily visible as $\e$
increases.  Since electronic computers became easily accessible it is
easy for everybody to observe personally on computer screens the very
complex drawings generated by such motions (as {\it seen} by Poincar\'e,
Birkhoff, Hopf \etc, {\it without} using a computer).

Furthermore quasi periodic motions although being, at least for $\e$
small, very common and almost dense in phase space probably do not
constitute an obstacle to the fact that the not quasi periodic motions
evolve very far, in the long run, from the points visited by the quasi
periodic motions to which the initial data were close. This is the
phenomenon of {\it Arnold's diffusion} of which there exist quite a
few examples: it is a phenomenon of wide interest.  For example, if in
the theory of the solar system diffusion was possible it would be
conceivable the occurrence of important variations of quantities such as
the radii of the orbits of the planets, with obvious (dramatic)
consequences on the stability of the solar system.

In this last question the true problem is the evaluation of the time
scale on which the diffusion in phase space could be observable.  In
systems simpler than the solar system (to which, strictly speaking,
Kolmogorov's theorem does not directly apply, for some reasons that we
shall not attempt to analyze here,\cite{[Ga84]} Sec. 5.10) one thinks
that a sudden transition, as the intensity $\e$ of the perturbation
increases, is possible from a regime in which the diffusion times are
super astronomical in correspondence of the interesting values of the
parameters (\ie times of several orders of magnitude larger than the
age of the Universe) to a regime in which such times become so short
to be observable on human scales.  This is one of the central themes
of the present day research on the subject, \cite{[MM87]}.

\end{section}

\*
\begin{verse}
\0{\it I know that I am mortal and the creature of a day; but when I
search out the massed wheeling circles of the stars, my feet no longer
touch the Earth, but, side by side with Zeus himself, I take my fill
of ambrosia, the food of Gods}: quotation of Ptolemy borrowed
from\cite{[He91]}, p. lvii.
\end{verse}
\*
\*

\def\revtex{R\raise2pt\hbox{E}VT\lower2pt\hbox{E}X}
\0(slightly corrupted) \revtex
\*\*

\baselineskip=10pt
\begin{verse}
\0{\it Giovanni Gallavotti\\
\0Dipartimento di Fisica\\
\0Universit\'a di Roma ``La Sapienza''\\
\0P.le Moro 2, 00185 Roma, Italy}\\
\*
\0email:\\
\0{\tt giovanni@ipparco.roma1.infn.it}\\
\0tel. +39-06-49914370, \\
\0fax. +39-06-4957697\\
\0web: {\tt http://ipparco.roma1.infn.it}
\end{verse}
\*


\begin{references}
\bibitem{nota1-1} \ie no linear combination of them with rational
coefficients can vanish unless all coefficients vanish.
\bibitem{nota2} Note that, from Newtonian mechanics and as discussed
below, the motions of the $8$ classical planets (the Fixed Stars Sky,
Sun, Mercury, Venus, Moon, Mars, Jupiter, Saturn not counting the
Earth (whose rigid motions are described by those of the Fixed Stars),
or alternatively not counting the Sun and the Fixed Stars but
regarding the Earth as having $6$ degrees of freedom, requires a
maximum of $24=3\times 8$ independent ``fundamental'' frequencies
namely three for each planet: so that both in Ptolemy and in
Copernicus there must be epicycles rotating at speeds multiple of the
fundamental frequencies or at least at speeds which are linear
combinations with integer coefficients of the fundamental frequencies:
hence the $43$ frequencies cannot be rationally independent of each
other; see for instance the figure on Copernicus' Moon and note that
the ``number of epicycles'' in Ptolemy's theory could be counted
differently.
\end{references}
\end{document}